# A high performance Nb nano-SQUID with a three-dimensional structure

*Lei Chen[1]\*, Hao Wang[1,2], Xiaoyu Liu[1], Long Wu[1,2], Zhen Wang[1,2,3]\**

\*Corresponding author:

Lei Chen (email: leichen@mail.sim.ac.cn; fax:86-21-62127493)

Zhen Wang (email: zwang@mail.sim.ac.cn; fax:86-21-62127493)

1. Center for Excellence in Superconducting Electronics, State Key Laboratory of Functional Material for Informatics, Shanghai Institute of Microsystem and Information Technology, Chinese Academy of Sciences, Shanghai 200050 China

2. University of Chinese Academy of Sciences, Beijing 100049, China

3. Shanghai Tech University, Shanghai 200031, China

## ABSTRACT

A superconducting quantum interference device (SQUID) miniaturized into nanoscale is promising in the inductive detection of a single electron spin. A nano-SQUID with a strong spin coupling coefficient, a low flux noise, and a wide working magnetic field range is highly desired in a single spin resonance measurement. Nano-SQUIDs with Dayem-bridge junctions excel in a high working field range and in the direct coupling from spins to the bridge. However, the common planar structure of nano-SQUIDs is known for problems such as a shallow flux modulation depth and a troublesome hysteresis in current-voltage curves. Here, we developed a fabrication process for creating three-dimensional (3-D) niobium (Nb) nano-SQUIDs with nano-bridge junctions that can be tuned independently. Characterization of the device shows up to 45.9 % modulation depth with a reversible current-voltage curve. Owning to the large modulation depth, the measured flux noise is as low as 0.34 $\mu\Phi_0/\mathrm{Hz}^{1/2}$. The working field range of the SQUID is greater than 0.5 T parallel to the SQUID plane. We believe that 3-D Nb nano-SQUIDs provide a promising step toward effective single-spin inductive detection.

The superconducting quantum interference device (SQUID) is one of the most sensitive magnetic flux detectors available. Moreover, by decreasing the size of the SQUID loop, the miniaturized SQUID also becomes a super-sensitive spin sensor.[1-3] Recently, the on-tip Pb nano-SQUID has demonstrated spin sensitivity below a single Bohr magneton.[4] The number is comparable to that of other supersensitive spin detection methods like the magnetic resonance force microscopy [5-7] and the NV center sensor found in diamonds [8-10]. Hence, it is likely that such a sensor would also prove very useful in nanoscale on-chip electron spin resonance (ESR) or nuclear magnetic resonance (NMR) spectroscopy.

The application of miniaturized SQUIDs pioneered by Wernsdorfer *et. al.* enabled important discoveries in the field of nano-magnetism.[1] Recently, other notable applications have also been demonstrated, such as the use of a high resolution scanning SQUID microscope,[4, 11-13] the displacement measurement of a nano-mechanic oscillator,[14, 15] and the inductive transition edge sensor.[16] Numerous nano-SQUIDs have been developed in recent years for a variety of purposes, such as those with Dayem bridge junctions, [12, 17-24] tri-layer junctions, [25-28] YBCO boundary junctions combined with nano-constrictions,[29, 30] and the one with SNS junctions based on proximity effect [31, 32]. Nevertheless, the most commonly used nano-SQUID remains the one with the nanoscale Dayem-bridge junctions; this is not only because of their compatibility with the high magnetic field,[18, 33] but also because the nano-bridge geometrically provides ideal near-field coupling geometry to an external spin. [34]

Unfortunately, the nano-SQUID with Dayem-bridge junctions is known for problems such as its shallow flux modulation depth and hysteresis in the current-voltage (*I-V*) curve.[1, 3, 22, 35-37] The shallow flux modulations associated with these junctions limits the flux noise of the device. At the same time, *I-V* hysteresis prevents the nano-SQUID from using standard SQUID readout electronics without special shunting.[38, 39] Recently, Vijay *et. al.*[40] developed the aluminum(Al) 3-D nano-SQUID with a layer of the superconducting loop thicker than the nano-bridge junctions via the shadow evaporation technique. It demonstrates a large flux modulation combined with an ultra-low flux noise,[40-42] and uses the long coherence length of the Al film and its shadow evaporation techniques to its advantage. Unfortunately, other films like niobium (Nb) associated with the higher critical field and critical temperature neither has such a long coherence length nor fits for the shadow evaporation techniques.

Here, we developed a fabrication process for a 3-D Nb nanoSQUID with tunable nano-bridge junctions. The fabricated devices demonstrate a large (up to 45.9 %) flux modulation depth, which is considerably greater than most planar Nb nano-SQUIDs (5-15% on average).[1, 22, 28, 43] Furthermore, the *I-V* curves of the 3-D structures are mitigated into a thermal regime [44] in which hysteresis disappears without any special shunting, and the readout via standard SQUID electronics becomes possible. The nano-SQUID made with Nb tolerates a working magnetic field up to 0.50 T parallel to the SQUID plane, which is sufficient to perform an X band (10-GHz) electron spin resonance (ESR) measurement. Owing to the large flux modulation, the intrinsic flux noise of 0.34 $\mu\Phi_0/Hz^{1/2}$ of the nanoSQUID is able to be measured by a simple low-noise room-temperature amplifier. Therefore, we consider that such a device is very promising in developing the on-chip detection for a small spin ensemble or even for single spin detection.

In contrast to a planar structure, a 3-D nanoSQUID is defined by a specific feature: that the layer of the SQUID loop is considerably thicker than that of two parallel nano-bridge junctions. In this way, not only is the inductance $L$ of the SQUID decreased, but the non-linearity of the current-phase relation of the nano-bridge junctions is also greatly improved. By using a specific lift-off technique,[45] as shown in Figure 1(a), an only 16-nm wide vertical insulating slit can be embedded in the middle of a thick superconducting film as shown in Figure 1(b). A nano-SQUID is formed by setting two parallel thin nano-bridges across the insulating slit on top of the thick niobium film. In principle, it is the width of the slit and the separation between the two nano-bridges that, taken together, determine the size of the nano-SQUID.

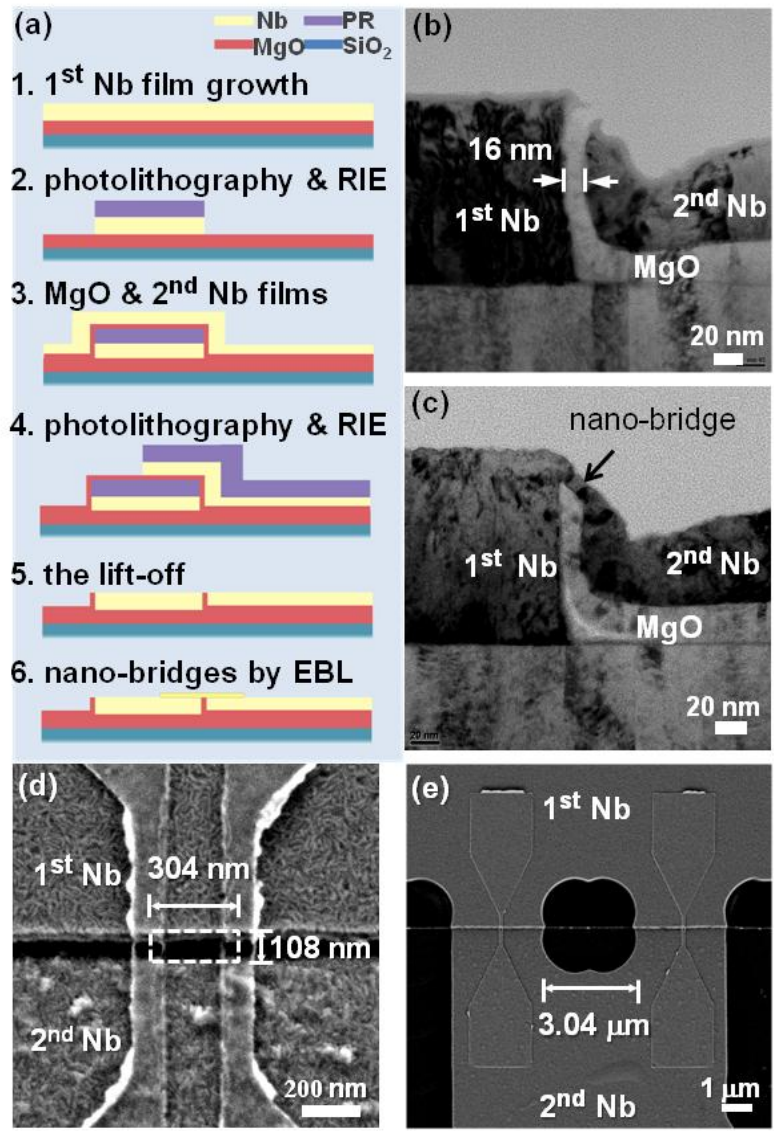

Figure 1 (a) the main steps of the 3-D Nb nano-SQUIDs fabrication process. (b) The transmission electron microscopy (TEM) picture of the cross-section of the insulating slit as in step V. (c) The TEM picture of the cross-section of the insulating slit with the nano-bridge on top as in step VI. (d) The top view of the nano-SQUID with two nano-bridge junctions across the insulation slit. (e) The top view of the SQUID with 3-μm square hole cut between the two nano-bridge junctions for an easier flux bias.

The fabrication process is illustrated in Figure 1(a) as the following. Step 1: on top of a 200-nm MgO film, a 130 nm thick Nb film was deposited via DC magnetron sputtering. Step 2: the Nb film was patterned by using UV photolithography followed by $CF^4$ reactive-ion etching (RIE). Afterwards, the photo-resist was kept on the sample on purpose. Step 3: another insulating MgO film was grown of 30 nm thickness, upon which a second Nb film of 105 nm thickness was deposited. Step 4: the second Nb film was again patterned by UV Photolithography combined with $CF^4$ RIE. Step 5: the photo-resist, together with the film on top of it, was lift-off by soaking the whole chip in an acetone solution. Step 6: above the Nb film, two parallel Nb nano-bridges—with a width of 100 nm and a thickness of 15 nm—were patterned across the MgO slit using electron-beam lithography (EBL). As shown in Figure 1(d), the fabricated nano-SQUID is of an geometric 108nm × 304nm superconducting loop. To help apply magnetic flux, we etched a 3-μm × 3-μm square hole in between the two nano-bridges, as shown in Figure 1(e).

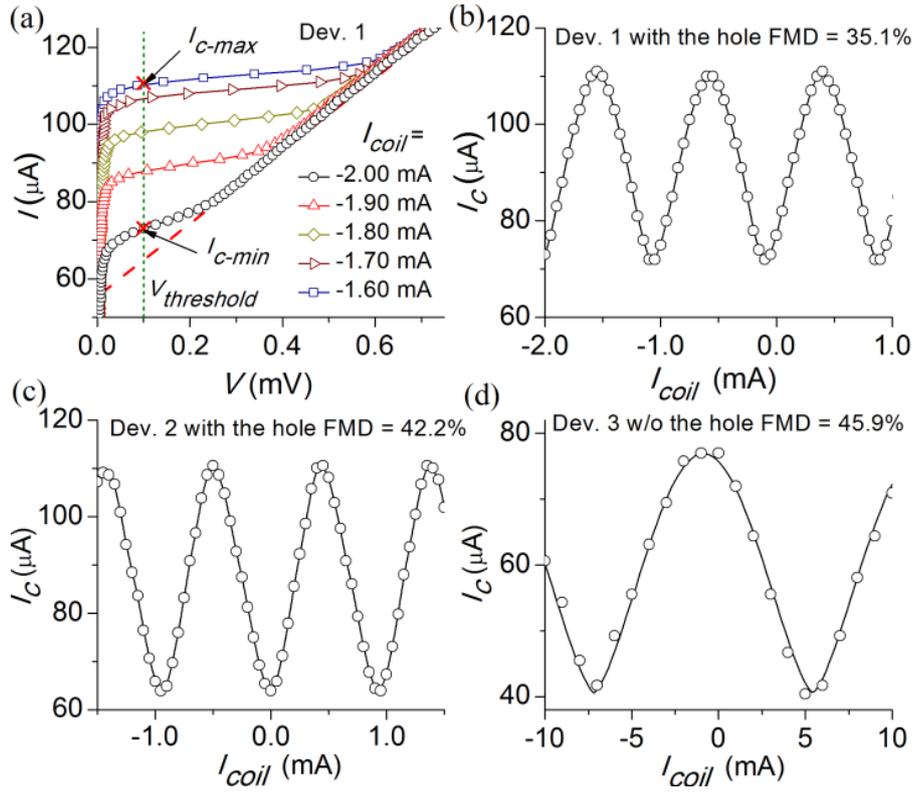

Figure 2: Electrical transport characterization at 4.20 K of two devices with a 3-µm square hole in the middle as in Figure 1(e), namely, Dev. 1 and Dev. 2, and one device without a hole in the middle as in Figure 1(d), namely Dev. 3. (a) the current (I)-voltage (V) curve of Dev. 1. (b)-(d) the Ic-flux modulation curves of Dev. 1, Dev. 2 and Dev. 3 respectively.

The current $I$ versus voltage $V$ curves of the nano-SQUID were measured in liquid helium at temperature $T$ = 4.20 K. Above the device, there was placed a small hand-wound superconducting coil that can apply the magnetic field perpendicular to the SQUID plane $B_\perp$ by applying current through the coil $I_{coil}$. In Figure 2(a), plots of the $I$-$V$ curves of Dev. 1 at various $I_{coil}$ are shown. The critical current $I_c$ is recorded at the place where it develops 0.1 mV voltage across the SQUID. In Figure 2(b)-(d), the magnetic flux modulation of $I_c$ is monitored by sweeping the $I_{coil}$ for Dev. 1, Dev. 2 and Dev. 3; a flux modulation depth $FMD$ of 35.1%, 42.2%, and 45.9% was observed respectively. Here, the $FMD$ is defined as $FMD = (I_{c\text{-}max}\text{-}I_{c\text{-}min})/I_{c\text{-}max}$, where $I_{c\text{-}max}$ and $I_{c\text{-}min}$ is the critical current at the constructive and destructive quantum interference, respectively. Here, Dev. 1 and 2 are from the same batch that had a 3-µm × 3-µm square hole etched between the two nano-bridges, as shown in Figure 1(e). Dev. 3 is the nano-SQUID that had two parallel nano-bridges across the insulating slit without any etched hole, as shown in Figure 1(d). Since the nano-bridge junctions and the superconducting loop are shaped in separate steps, the $I_c$ and $L$ can tuned independently to maximize the FMD of the nano-SQUID.

It is noteworthy that the $I$-$V$ curves in Figure 2(a) are completely reversible. Unlike with conventional SQUIDs, the problematic $I$-$V$ hysteresis of nano-bridge junction SQUIDs is primarily induced by thermal heating. In 2-D structures, it has been found that there is a small thermal region where $I_c$ is sufficiently small to just switch the nano-bridges and the $I$-$V$ curves are reversible.[44] However, in the 3-D structure, the superconducting bank is considerably thicker than the nano-bridge junctions and remains strongly superconducting even after the nano-bridges are switched. The thermal heating induced by switching the nano-bridges is relatively minor and dissipates easily.

As the *I-V* curves of the nano-SQUID are reversible, it is possible to use a standard SQUID readout circuit to characterize its magnetic flux noise. Here, we connected the output of the SQUID to a customized 30-dB low-noise room-temperature amplifier ($\sim$2 nV/Hz$^{1/2}$) followed by a high speed analog data acquisition card. The Dev. 2 was fixed at the red sensitive point, as in the upper-right insert shown in Figure 3, by sending a bias current $I_{bias}$ = 65 μA and by applying $I_{coil}$ = 119.5μA to the flux coil at the same time. The noise density spectrum in Figure 3 was acquired by the fast Fourier transform of the voltage data from the acquisition card. Owning to the large $dV/d\Phi$ = 17.7 mV/$\Phi_0$ resulting from the large flux modulation depth, the flux-noise induced voltage-noise is greater than the amplifier noise. The measured white flux noise of Dev. 2 is as low as 0.34 μ$\Phi_0$/Hz$^{1/2}$.

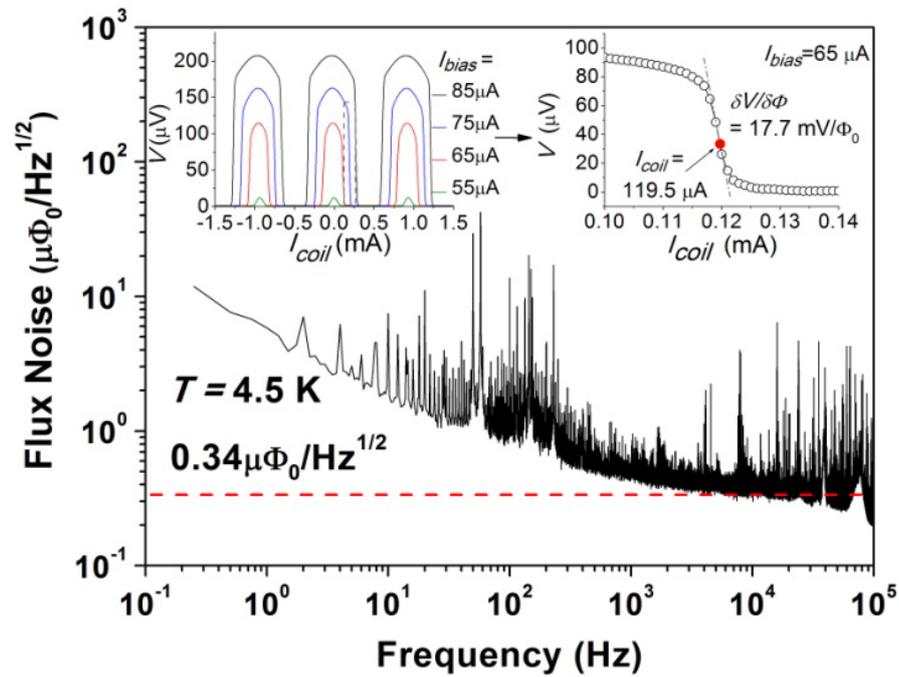

Figure 3 The magnetic flux noise density spectrum of Dev. 2 at T = 4.5 K. The upper-left insert plots the voltage across the SQUID as a function of the coil current at various current bias points. The upper-right insert zooms in to the sensitive working point at which the flux noise density spectrum was measured.

In an on-chip spin resonance experiment, the nano-SQUID is coupled with the spins directly under the magnetic field. Therefore, it is important that the sensor can endure a relatively high magnetic field. To characterize its working field range, the 3D nano-SQUID Dev. 4 was mounted on a rotatable puck in the Physics Property Measurement System (PPMS), and the puck was rotated to an angle at which the SQUID plane was parallel to the magnetic field. Dev. 4 is of the same structure as Dev. 2. As a hand-wound coil is not possible when using a PPMS puck, the magnetic flux is applied by the $I_{coil}$ through the line joint with SQUID washer directly, as in the cartoon insert shown in Figure 4. In Figure 4, the $I_{c-max}$ and $I_{c-min}$ of the $I_c$ flux modulation curve were plotted as a function of the applied parallel magnetic field $B_{//}$. Because if fluctuations in the magnetic field and the rotation angle, there is an uncertain out-of-plane field $B_{\perp}$ component that deviated the measured $I_{c-max}$ and $I_{c-min}$ from the parabola fitting lines. Fortunately, the out-of-plane field is quiet enough to allow us to observe the modulation curves, as shown in the insert in Figure 4. A clean modulation curve is observed at 0.5 T, meaning that the nano-SQUID is functional at least at 0.5 T; this corresponds to the field of an X-band (10 GHz) ESR requirement.

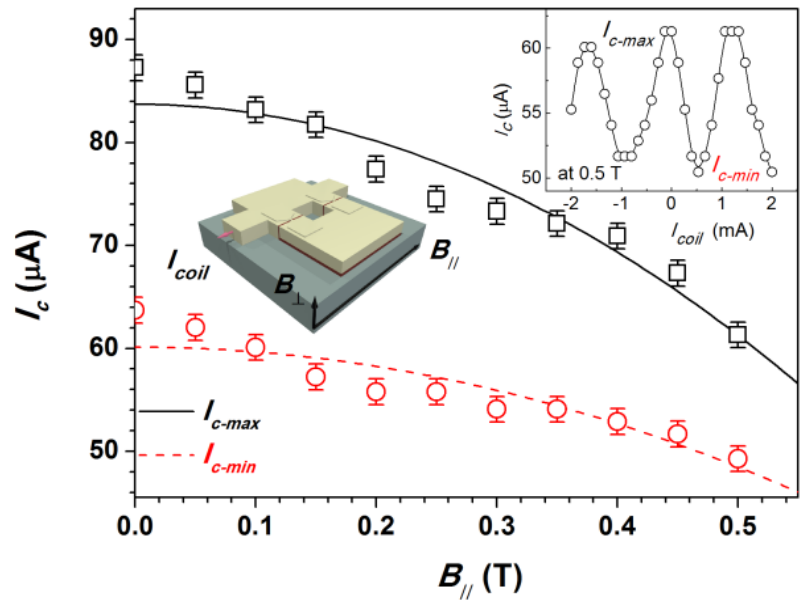

Figure 4 The $I_{c\text{-}max}$ and $I_{c\text{-}min}$ as a function of the in-plane field $B_{//}$ of Dev. 4. $I_{c\text{-}max}$ and $I_{c\text{-}min}$ are the critical current of the constructive and destructive interference respectively. The insert shows the flux modulation curve at $B_{//} = 0.5$ T.

The nano-SQUID shows a zero-voltage interception at $I_{c\text{-}max}/2$ in the I-V curves, as shown in figure 2(a). The same behavior was described in quantum phase-slip (QPS) center characterization through Skocpol-Beasley-Tinkham model.[46] The interception was a time-averaged super-current in the Josephson cycle. The existence of the QPS indicates the effective length of our constrictions remains greater than the superconducting coherence length $\xi_0$. Although as shown in figure 1(b) the insulating slit is only 16 nm wide, the edge of one side of the bank is not completely flat due the shadow effect during the sputtering. The roughness elongate the effective length of Nb nano-bridge. In the worst case, we consider the length of nano-bridges to be 108-nm as shown in Figure. 1(d). By further optimizing the process and reducing the effective length $L_{eff}$ of the nano-bridge, the zero-voltage interception will be minimized for the *I-V* curve at the destructive interference and the FMD can be potentially further increased [47].

The minimal detectable spin number for a nano-SQUID in a spin resonance experiment can be written as $S_N = \frac{\alpha S_{\Phi}}{\mu_0 p \mu_B}$. Here, $\alpha$ is the coupling factor between the spin ensemble and the nanoSQUID, $S_{\Phi}$ is the flux noise of the SQUID, and $p$ is the spin polarization ratio at thermal equilibrium; for the latter, $p = \tanh\left(-\frac{g\mu_B B}{2k_B T}\right)$ where $\mu_0$ is the permeability constant, $\mu_B$ is the Bohr magneton, $g$ is the g-factor, $B$ is the magnetic field, $k_B$ is the Boltzmann constant, and $T$ is the temperature. Therefore, a stronger coupling $\alpha$, a low flux noise $S_{\Phi}$, and a high magnetic field $B$ are highly desired as part of a nano-SQUID detected on-chip spin resonance experiment. [7, 48] In a simple assumption that a square loop made of infinitely narrow wires has a side length $l$, $\alpha = \sqrt{2}\pi l/4$.[3] As the Dev.2 has a square middle hole with a side length of 3 μm, the estimated spin sensitivity is 199 $\mu_B/Hz^{1/2}$. If the side length of a SQUID loop is further minimized toward the EBL limit by further refinement of the process, assuming 10 nm, the achievement of spin sensitivity of 0.66 $\mu_B/Hz^{1/2}$ is possible using the 3-D nano-SQUID. Since our devices are made of films instead of infinitely narrow wires, the simple estimation here is optimistic. A finite-element simulation based on the device's shape and size will be required to obtain a more accurate spin sensitivity.[49] On the other hand, it has also been proposed that the spin sensitivity can be also increased by a near-field coupling scheme in which the spin is coupled to the nano-constrictions directly. [34]

In conclusion, we developed a fabrication process for creating a 3-D Nb nano-SQUID with tunable nano-bridge junctions. The fabricated SQUID shows a flux modulation depth up to 45.9 %, which is considerably greater than its 2-D counterpart. It also shows a reversible *I-V* curve that permits a simple standard SQUID readout. Benefitting from the large flux modulation depth and reversible I-V curve, the intrinsic flux noise acquired by a simple room-temperature amplifier is as low as 0.34 $\mu\Phi_0/Hz^{1/2}$. Moreover, the flux modulation as a function of the in-plane magnetic field shows that the device works at up to 0.5 T, which is more than sufficient for X-band ESR spectroscopy. In addition, the zero voltage interception at $I_{c\text{-}max}/2$ in the *I-V* curves indicate that the effective length of the nano-bridge junction is still longer than its superconducting coherence length. Further refinement of the fabrication process may make the flux modulation even deeper, and a single electron spin on-chip detection is possible.

**Corresponding Author:**

*Lei Chen: leichen@mail.sim.ac.cn

*Zhen Wang: zwang@mail.sim.ac.cn

**Author Contributions**

LC and ZW planned the research. LC, HW and LW performed the experiments and collected the data. XL did the electron-beam lithography. LC and HW analyzed the data. LC and ZW wrote the paper. All authors approved the final version of the manuscripts.

**ACKNOWLEDGEMENT**

We acknowledge support from the Strategic Priority Research program of the Chinese Academy of Sciences (Grant No. XDB04000000), as well as funding from the National Science Foundation of China (Grant No. 61306151) and from the State Key Laboratory of Functional Materials for Informatics (Grant No. SKLFMI201504).

TABLE OF CONTENTS GRAPHIC:

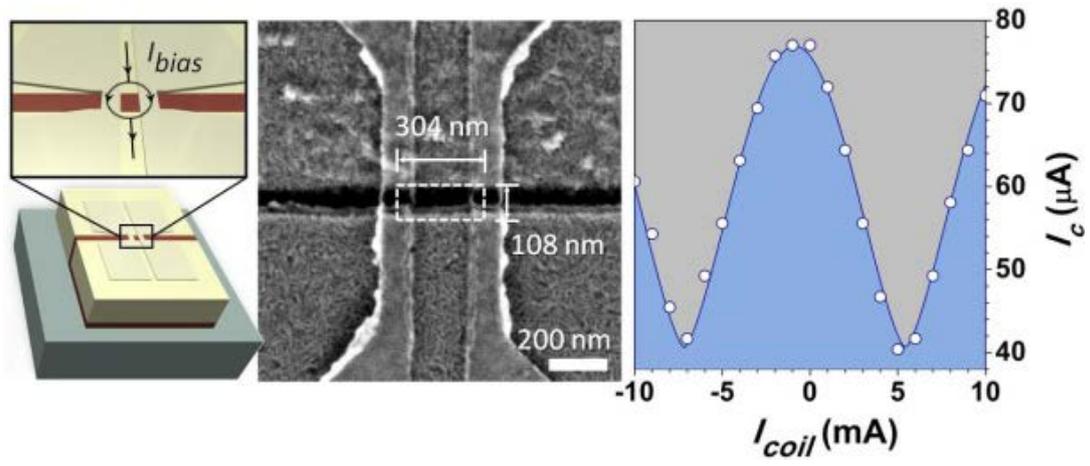